\documentclass[a4paper]{article}
\usepackage{amsthm,amssymb,amsmath} 
\usepackage{hyperref} 
\usepackage{subfigure} 
\usepackage{graphicx} 

\usepackage[top=1.2in, bottom=1.2in, left=1.2in, right=1.2in]{geometry}
\usepackage{setspace}
\usepackage{natbib}
\renewcommand{\times}{~}
\newcommand{\set}[1]{\mathcal{#1}}

\newcommand{\x}{\mathbf{x}}

\newcommand{\mTitle}{An Integer Linear Programming Model for the Radiotherapy Treatment Scheduling Problem}
\newcommand{\mAuthor}{Edmund K. Burke\asep Pedro Leite-Rocha\asep Sanja Petrovic}
\newcommand{\mKeywords}{Health care\ksep Applications\ksep Integer Programming\ksep Patient scheduling}
\newcommand{\mSubject}{Radiotherapy Treatment Scheduling}

\newcommand{\asep}{, }
\newcommand{\ksep}{; }
\begin{document}
\renewcommand{\asep}{\and }
\renewcommand{\ksep}{\and }
\title{\mTitle}
\author{\mAuthor}
\date{\emph{Automated Scheduling, Optimisation and Planning (ASAP) research group
  \\School of Computer Science, University of Nottingham, Nottingham, UK}}
\maketitle

\begin{abstract}
  Radiotherapy represents an important phase of treatment for a large number of cancer patients. It is essential that
  resources used to deliver this treatment are employed effectively. This paper presents a new integer linear
  programming model for real-world radiotherapy treatment scheduling and analyses the effectiveness of using this model
  on a daily basis in a hospital. Experiments are conducted varying the days on which schedules can be created. Results
  obtained using real-world data from the Nottingham University Hospitals NHS Trust, UK, are presented and show how the
  proposed model can be used with different policies in order to achieve good quality schedules.
\end{abstract}

\section{Introduction}

The number of cancer cases in the United Kingdom has greatly increased in the last few decades. Approximately 200,000
new cases of cancer are discovered in England per year, causing 120,000 deaths \citep{nhs2000}. Although treatment
has improved recently, there is still much room for improvement \citep{nhs2004}. Several audits conducted in the UK
show that the waiting times for cancer treatment are not yet satisfactory
\citep{rcr1998,spurgeon2000,ash2004,summers2006,drinkwater2008}. Even though cancer care has improved recently within
the United Kingdom, radiotherapy capacity is still an important factor that has not received the adequate amount of
attention \citep{dodwell2006}. New targets set by the \cite{nhs2007}, in a program devised to enhance the effectiveness
of cancer treatment in the UK, make radiotherapy scheduling a very important problem.

This paper deals with a real-world radiotherapy treatment scheduling problem present in the Nottingham University
Hospitals NHS Trust, UK. The main aim is to design, implement and validate a scheduler for the linear accelerators
(hereafter referred to as linacs) used to deliver radiotherapy treatment. The purpose of the scheduler is to assist the
radiographer at the end of each day to create a schedule for the patients who are available to be scheduled.

There are some similarities between radiotherapy treatment scheduling and other appointment scheduling problems. Such
problems have the objective of optimising some quality of service measure, usually related to the waiting time of the
patient before being seen by a doctor, and deal with a stochastic arrival of patients. However, there are some key
differences as well. In other appointment scheduling problems, the objective is, usually, to schedule a single doctor
appointment for a patient, which often has a stochastic duration, and the patient often must be informed of the time of
the appointment immediately at the time of request. Those do not happen in radiotherapy treatment scheduling. The
objective here is to schedule a given number of appointments of deterministic duration with specific time intervals
between them for each patient. In addition, the scheduling of patients is usually done in batches once per day, in
order to find a better schedule.

As far as the authors are aware, few papers in the scientific literature deal with the problem of scheduling
radiotherapy treatments. \cite{petrovic2009} develop a genetic algorithm approach which considers both pre-radiotherapy
and radiotherapy treatment. \cite{kapamara2006} give a review of the radiotherapy patient scheduling (both
pre-treatment and treatment) problem, concluding that this problem is similar to a dynamic stochastic job-shop problem.

Mathematical and simulation models are commonly used to approach the problem or other similar problems.
\cite{conforti2008} define mathematical models for the scheduling of radiotherapy treatment. The objective in their
proposed model is to schedule as many patients as possible in a short period of time (e.g. one week). They consider a
block system, where a workday is split into a fixed number of time blocks/slots. In a successive paper, the same
authors extend the model to take patient availability into account, and run more extensive experiments
\citep{conforti2009} with real world data. \cite{conforti2010} then consider a non-block system, where the session time
may vary from one session to another. They observe that uniform appointment blocks do not represent real workload
properly, since the treatments can take either more or less time than the chosen block time. However, the models do not
consider all the constraints present in real-world radiotherapy scheduling, such as linac eligibility, treatments which
are not held on consecutive days, release dates different from the booking requests and patients who require multiple
sessions per day.

\cite{lev1974} develop a discrete event simulation model of patient flow in a diagnostic radiology department. This
model can be used to evaluate algorithms to be used for scheduling patients in such an environment. \cite{proctor2007}
propose a simulation model for a radiotherapy centre in the UK. The authors analyse two strategies to improve cancer
waiting times: 1) acquiring more equipment, such as a simulator and/or linac and 2) changing the working policy, such
as not requiring that radiographers treat the same patients for all sessions, extend working hours, etc.

The authors' previous research was concerned with constructive approaches to radiotherapy scheduling
\citep{petrovic2006,petrovic2008,petrovic2008a}. These methods vary some scheduling parameters, such as how
frequently to create schedules, and investigate the effect of changing the values of these parameters on the
performance of the algorithm. An algorithm based on the meta-heuristic GRASP was also developed to try to improve the
schedule generated by these constructive approaches.

The main contribution of this paper lies in the introduction of a new Integer Linear Programing (ILP) model for
scheduling radiotherapy treatments. Additional data from the hospital is gathered in order to better understand and
represent the radiotherapy scheduling problem. Experiments are conducted to evaluate the model in a myopic approach
varying the days on which schedules can be created, where no attempt is made to predict the patients who will arrive in
the near future.

This paper is organised as follows: \autoref{sec:problemdefinition} introduces radiotherapy treatment and describes
the procedure in place in the Nottingham University Hospitals NHS Trust, UK. The mathematical formulation of the
radiotherapy treatment scheduling problem is presented. In \autoref{sec:experiments}, experimental results are
presented. \autoref{sec:conclusion} gives conclusions about the experiments and future directions in this research.

\section{The Radiotherapy Treatment Scheduling Problem}
\label{sec:problemdefinition}

The radiotherapy treatment scheduling problem can be defined as the problem of scheduling a given number of
radiotherapy treatment sessions on linear accelerator machines. It is considered as a daily problem in which a number
of patients to be scheduled arrive at a radiotherapy centre. At the end of each day, the radiographer creates a
schedule for patients who arrived that day on a booking system which is partially booked with patients from previous
days.

Patients are grouped into different categories based on their waiting list status, which is classified as emergency,
urgent or routine. The status of a patient is decided by considering the site and the level of advancement of the
tumour. Patients are also grouped according to their treatment intent as radical (with the intent to cure) and
palliative (with the intent to alleviate symptoms and improve a patient's quality of life). This classification is also
used in the most recent audits by the Royal College of Radiologists \citep{summers2006,drinkwater2008}.

Each patient requires one or more types of radiation from the linacs, where the available radiation types are high
energy photon, low energy photon and electron. Since not all linacs can emit all types of radiation, this imposes a
linac eligibility constraint.

Patients can require more than one radiotherapy session, and the session duration can differ from one patient to
another, or even amongst the sessions of the same patient. Commonly, the first session of each patient is longer due to
validation and verification procedures.

Each linac can attend only one patient at a time. The capacity of each linac, given by the number of working hours of
the hospital staff, must not be exceeded on any given day.

Patients must undertake a set of preparatory steps before starting treatment, which are conveniently referred to as the
pre-treatment stage. Each patient can start treatment on or after the date when their pre-treatment finishes. This date
is referred to as the release date.

Patients may require 1, 2, 3 or 5 session days per week, where a session day is simply a day when the patient is
required to go to the radiotherapy centre to receive one or more fractions of the treatment. Patients who can be
treated on weekends can require up to 7 session days per week. Sessions must be scheduled with a strict number of days
between them, as such:
\begin{itemize}
  \item patients who have 1 session day per week must have all sessions on the same day of the week for consecutive
  weeks,
  \item patients who have 2 session days per week must be scheduled either on Mondays and Thursdays consecutively or on
  Tuesdays and Fridays consecutively,
  \item patients with 3 session days per week must be scheduled on Mondays, Wednesdays and Fridays consecutively,
  \item patients with 5 session days per week must have them on consecutive days excluding weekends,
  \item and patients with 7 session days per week must have them on consecutive days including weekends.
\end{itemize}

Some patients must have a minimum number of sessions before the first weekend in order to prevent the tumour from
growing back after the first sessions. For example, some palliative patients must have at least 2 sessions before the
first weekend, and therefore, cannot start their treatment on a Friday. Some patients with 5 or less sessions are
required to have them all on the same week on consecutive days, without interruptions.

\begin{figure}[tb]
  \centering
  \includegraphics[width=.9\textwidth]{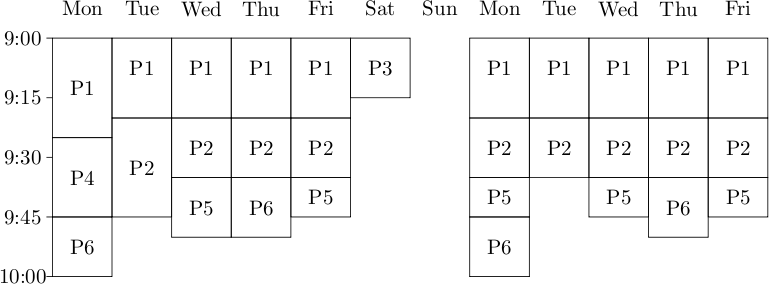}
  \caption{An example of a typical schedule where the duration of the first session of each patient is slightly longer
  than the others. Patient P1 has 10 sessions, 5 per week; patient P2 has 9 sessions, 5 per week; patient P3 is a
  emergency patient with 1 session on Saturday; patient P4 has 1 session; patient P5 has 5 sessions, 3 per week; and
  patient P6 has 4 sessions, 2 per week.}
  \label{fig:schedule}
\end{figure}

The majority of patients have only 1 session per day. The exceptions are CHART patients (Continuous Hyper-fractionated
Accelerated Radiotherapy Treatment), who require 3 fractions per day for 12 consecutive days with treatment starting on
a Monday. In addition, the presence of the doctor is required for the first session of some patients. Since each doctor
is available in the radiotherapy centre on only a few days of the week, this imposes one more eligibility constraint.
\autoref{fig:schedule} shows an example of a schedule for one linac where the opening times are set as from 9:00 to
10:00 for simplicity.

\begin{figure}[tb]
  \centering
  \includegraphics[width=.9\textwidth]{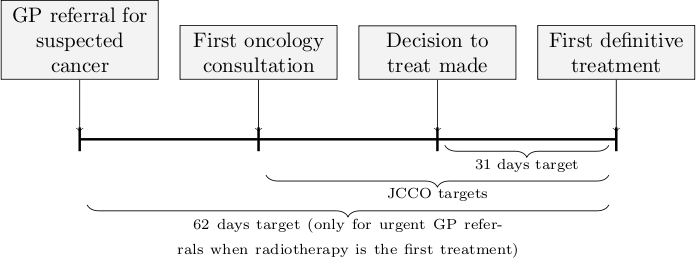}
%
%
%
%
  \caption{Time-line and Department of Health targets of patients diagnosed with cancer.}
  \label{fig:timeline}
\end{figure}

Three target dates are set for each patient. The first is established by the \cite{nhs2004}. It states that each
patient must start their treatment no later than 62 days from the date upon which they are referred to an oncologist by
their general practitioner (GP) and no later than 31 days from the date when the decision to treat with radiotherapy
was made. This is referred to as the \emph{breach date}. The UK Cancer Network evaluates each radiotherapy centre
according to the number of patients that breach this target, thus minimising this number is the primary objective in
this research. The Department of Health targets are illustrated in \autoref{fig:timeline} \citep{nhs2005}.

The other two target dates have been established by the \cite{jcco1993}. They determine the good practice and the
maximum acceptable waiting times from the date the patient is first seen for suspected cancers to the first session of
treatment for each category of patients. \autoref{tab:jccotargets} shows the JCCO waiting time targets which have
been adjusted to the nomenclature used in this work. The JCCO targets are acknowledged by the majority of radiotherapy
centres in the UK \citep{ash2004,summers2006} and considered as a secondary objective in this research. It should
also be noted that the targets used for emergency patients are 24/48 hours for the good practice/maximum acceptable
waiting times, while the targets for urgent and routine patients depend on treatment intent. This better reflects the
nomenclature currently in use in hospitals and has been suggested by \cite{drinkwater2008}.

\begin{table*}[tb]
  \centering
  \begin{tabular}{l||c|c||c|c||c|c}
    \hline
                       & \multicolumn{2}{c||}{emergency} & \multicolumn{2}{c||}{urgent} & \multicolumn{2}{c}{routine} \\
    \cline{2-7}
                       & palliative & radical            & palliative & radical         & palliative & radical        \\
    \hline \hline
    Good Practice      & 24 hours   & 24 hours           & 48 hours   & 2 weeks         & 48 hours   & 2 weeks        \\
    Maximum Acceptable & 48 hours   & 48 hours           &  2 weeks   & 4 weeks         &  2 weeks   & 4 weeks        \\
    \hline
  \end{tabular}
  \caption{Waiting time targets established by the JCCO adjusted to the nomenclature used in this work.}
  \label{tab:jccotargets}
\end{table*}

In addition to these three targets, the minimisation of waiting time from the decision to treat to the start of
treatment is considered. The hospital aims to minimise the waiting time while distributing it as evenly as possible
amongst patients. To measure it, the weighted sum of squared waiting times is calculated. This criterion can be
frequently seen in the literature for machine scheduling \citep{bagchi1987}, often used when large deviations of
completion time from the due date are undesirable.

To illustrate how the sum of squared waiting times can be applied to our problem and how it differs from other
frequently used criteria such as the sum of waiting times or the maximum waiting time, let us consider the following
example: on a given day, 3 patients arrive at the radiotherapy centre to be scheduled. Let us suppose that in one
possible schedule, patients 1, 2 and 3 have a waiting time of 1, 3 and 3, respectively, while in a second schedule, the
waiting times are 2, 2 and 3. The hospital would prefer the second schedule since it distributes the waiting time more
evenly among the patients. However, if either the sum of waiting times or the maximum waiting time are used, the value
of the objective function for each schedule will be the same for both schedules (7 and 3 respectively) making them
indistinguishable. If the sum of squared waiting times is used, the value of the objective function will be 19 and 17
for the first and second schedule respectively, enabling the algorithm to correctly choose the second schedule, which
would be preferred by the hospital.

A weight is assigned to each patient, which defines the relative importance of that patient respecting the JCCO
targets. As done in previous work \citep{petrovic2008}, the weights are set to 10 for emergency patients, 3 for urgent
and 1 for routine.

\section{Integer Linear Programming Model}
\label{sec:problemstatement}

The problem described can be formulated as an integer linear programming (ILP) model with the following input data:
\begin{itemize}
  \item $M$: number of linacs,
  \item $i$: index for linacs ($i = 1, \ldots, M$),
  \item $N$: number of patients available to be scheduled,
  \item $j$: index for patients available to be scheduled ($j = 1, \ldots, N$),
  \item $\set{M}_j$: set of machines which can emit the required radiation types for patient $j$ ($\set{M}_j \subseteq
  \{1, \ldots, M\},$ $\set{M}_j \ne \emptyset$),
  \item $\set{W}_j$: set containing the days of the week when patient $j$ is allowed to have his/her first session
  taking into consideration if the patient can receive treatment on weekends, the number of sessions a patient must
  have before the first weekend, and the days the doctor is present at the hospital, if the doctor's presence is
  required for the first treatment
  ($\set{W}_j \subseteq \{$Mon$, \ldots, $Sun$\}, \set{W}_j \ne \emptyset$),
  \item $w_j$: relative importance (weight) assigned to patient $j$,
  \item $b_j$: date when the booking request of patient $j$ is made,
  \item $r_j$: release date of patient $j$,
  \item $d^1_j$: breach date by which patient $j$ should start the treatment as established by the \cite{nhs2005},
  \item $d^2_j$: maximum acceptable date by which patient $j$ should start the treatment as established by the
  \cite{jcco1993},
  \item $d^3_j$: good practice date by which patient $j$ should start the treatment as established by the
  \cite{jcco1993},
  \item $T$: number of days in the scheduling horizon,
  \item $k$: index for days in the scheduling horizon ($k = 1, \ldots, T$),
  \item $q_k$: day of the week of day $k$ ($q_k \in \{$Mon$, \ldots, $Sun$\}$),
  \item $c_{ik}$ - available capacity of linac $i$ on day $k$ given in minutes,
  \item $S_j$: number of sessions required for patient $j$,
  \item $l$: index for sessions of patient $j$ ($l = 1, \ldots, S_j$),
  \item $p_{jl}$: duration of session $l$ of patient $j$ given in minutes,
  \item $u_{jkl}$: number of days patient $j$ must wait between sessions $l$ and $l + 1$ if session $l$ is scheduled on
  day $k$, or 0 if sessions $l$ and $l + 1$ are on the same day.
\end{itemize}

The model is composed of only one set of decision variables, defined as follows:
\[
  x_{ijkl} = \begin{cases}
    1 & \mbox{if session } l \mbox{ of patient } j \mbox{ is scheduled on day } k \mbox{ on linac } i,\\
    0 & \mbox{otherwise.}
  \end{cases}
\]

The first constraints are presented to ensure that sessions are not scheduled on any invalid machine or day.
Constraint (\ref{eq:eligibility1}) imposes that sessions of patient $j$ are not scheduled on machines that do not emit
the types of radiation required for patient $j$. Constraints (\ref{eq:eligibility2})-(\ref{eq:eligibility4}) ensure
that patients are not scheduled on invalid days. Constraint (\ref{eq:eligibility2}) imposes that the day of any session
of patient $j$ cannot be schedule before the release date, constraint (\ref{eq:eligibility3}) guarantees that the first
session of the patient is not on an invalid day of the week for that patient, and constraint (\ref{eq:eligibility4})
ensures that no session other than the first of each patient can take place on the first day of the scheduling horizon.
\begin{eqnarray}
  x_{ijkl} = 0 && i = 1, \ldots, M, i \notin \set{M}_j, j = 1, \ldots, N, k = 1, \ldots, T, l = 1, \ldots, S_j
  \label{eq:eligibility1}\\
  x_{ijkl} = 0 && i = 1, \ldots, M, j = 1, \ldots, N, k = 1, \ldots, r_j - 1, l = 1, \ldots, S_j
  \label{eq:eligibility2}\\
  x_{ijk1} = 0 && i = 1, \ldots, M, j = 1, \ldots, N, k = 1, \ldots, T, q_k \notin \set{W}_j
  \label{eq:eligibility3}\\
  x_{ij1l} = 0 && i = 1, \ldots, M, j = 1, \ldots, N, l = 2, \ldots, S_j
  \label{eq:eligibility4}
\end{eqnarray}

Each pair of sessions of the same patient must be scheduled $u_{jkl}$ days apart, depending on the day $k$ when session
$l$ is scheduled. To ensure that session $l+1$ is scheduled $u_{jkl}$ days after session $l$, constraint
(\ref{eq:sequence}) is included.
\begin{multline}
  x_{ijk'l'} = x_{ijkl} \qquad k' = k + u_{jkl}, l' = l + 1,\\
  i = 1, \ldots, M, j = 1, \ldots, N, k = 1, \ldots, T - u_{jkl}, l = 1, \ldots, S_j - 1,
  \label{eq:sequence}
\end{multline}

It is necessary to guarantee that all sessions are scheduled, and that each session is scheduled on exactly one day
and one linac. Constraint (\ref{eq:onelinac}) imposes this restriction.
\begin{equation}
  \sum_{i=1}^M \sum_{k=1}^T x_{ijkl} = 1 \qquad j = 1, \ldots, N, l = 1, \ldots, S_j
  \label{eq:onelinac}
\end{equation}

Finally, the available capacity on linacs must be respected. Constraint (\ref{eq:capacity}) ensures that the total
time used by sessions on day $k$ on linac $i$ does not exceed the linac capacity for that day.
\begin{equation}
  \sum_{j=1}^N \sum_{l=1}^{S_j} p_{jl} \times x_{ijkl} \leq c_{ik} \qquad i = 1, \ldots, M, k = 1, \ldots, T
  \label{eq:capacity}
\end{equation}


The criteria considered here are described below and are presented in order of their importance. This order has been
decided according to hospital staff preference.

\begin{itemize}
  \item Minimisation of the number of patients who miss the breach date:
  \begin{equation}
    f_1(\x) = \sum_{i=1}^M \sum_{j=1}^N \sum_{k=d^1_j+1}^T x_{ijk1},
    \label{eq:objbreach}
  \end{equation}
  \item Minimisation of the weighted number of patients who miss the JCCO maximum acceptable target:
  \begin{equation}
    f_2(\x) = \sum_{i=1}^M \sum_{j=1}^N \sum_{k=d^2_j+1}^T w_j \times x_{ijk1},
    \label{eq:jccomax}
  \end{equation}
  \item Minimisation of the weighted number of patients who miss the JCCO good practice target:
  \begin{equation}
    f_3(\x) = \sum_{i=1}^M \sum_{j=1}^N \sum_{k=d^3_j+1}^T w_j \times x_{ijk1},
    \label{eq:jccogood}
  \end{equation}
  \item Minimisation of the weighted average squared waiting times:
  \begin{equation}
    f_4(\x) = \sum_{i=1}^M \sum_{j=1}^N \sum_{k=b_j+1}^T (k - b_j)^2 \times w_j \times x_{ijk1}.
    \label{eq:waitingtimesqrd}
  \end{equation}
  It should be noted that, even though the squared waiting time is being calculated, no decision variables are in fact
  squared and the model remains linear.
\end{itemize}

In order to handle multiple objectives optimisation, a lexicographical ordering \citep{steuer1986,yu1989} is used. The
set of $Y$ objectives is indexed so that objective $m$ is more important than objective $m + 1$. A lexicographical
ordering preference is defined as follows: solution $\x^1$ is preferred to solution $\x^2$ iff $f_1(\x^1) < f_1(\x^2)$
or there is some $m \in \{2, \ldots, Y\}$ so that $f_m(\x^1) < f_m(\x^2)$ and $f_{m'}(\x^1) = f_{m'}(\x^2)$ for $m' =
1, \ldots, m - 1$.


\section{Experiments and Results}
\label{sec:experiments}

In order to evaluate the model, experiments are run simulating the everyday scheduling of a hospital. Each day, a
number of patients arrive in the radiotherapy department to be scheduled. At the end of the day, the radiographer
creates a schedule for the patients that arrived on that day.


Two sets of data were given to the authors by the Nottingham University Hospitals NHS Trust, UK. The first set contains
the waiting list status, intent, radiation and booking request of patients treated in a period of five years. The
second set contains all the attributes from each patient necessary to build a schedule of patients treated in a period
of one month. Similarly to previous work \citep{petrovic2008,petrovic2008a}, these two sets are combined to create 33
different data sets to be used in the experiments by the following procedure:
\begin{itemize}
  \item Select a random period of time of 18 months of length from the first data set.
  \item For each patient in that period, use the attributes of a random patient from the second data set with the same
  waiting list status, intent and radiation to fill in the missing attributes in the first data set.
\end{itemize}
A warm-up period of 6 months is used, where the patients who arrive in the first 6 months are used only to fill the
booking system.

It should be noted that some combinations of waiting list status, treatment intent and required radiation type are more
frequent than others in the data received, and some are not present. The proportion of each combination of these
attributes can be seen in \autoref{tab:patienttype}, where patients marked as requiring electron radiation may also
require low energy photon and patients marked as requiring high energy photon may also require electron and low energy
photon.

\begin{table}[tb]
  \centering
  \begin{tabular}{l|c|c||c|c||c|c}
    \hline
    Required & \multicolumn{2}{c||}{Emergency} & \multicolumn{2}{c||}{Urgent} & \multicolumn{2}{c}{Routine} \\
    \cline{2-7}
    Radiation Type      & Palliative & Radical & Palliative & Radical & Palliative & Radical\\
    \hline
    \hline
    High energy photon & 1.3 & --  & 17.1 & --  & --  & 20.5 \\
    Low energy photon  & 2.4 & --  & 14.4 & --  & 2.8 & 15.1 \\
    Electron           & --  & --  & 10.2 & --  & 1.5 & 14.6 \\
    \hline
  \end{tabular}
  \caption{Proportion (\%) of each combination of waiting list status, treatment intent and required radiation type}
  \label{tab:patienttype}
\end{table}

\begin{figure}[tb]
  \centering
  \includegraphics[width=.8\textwidth]{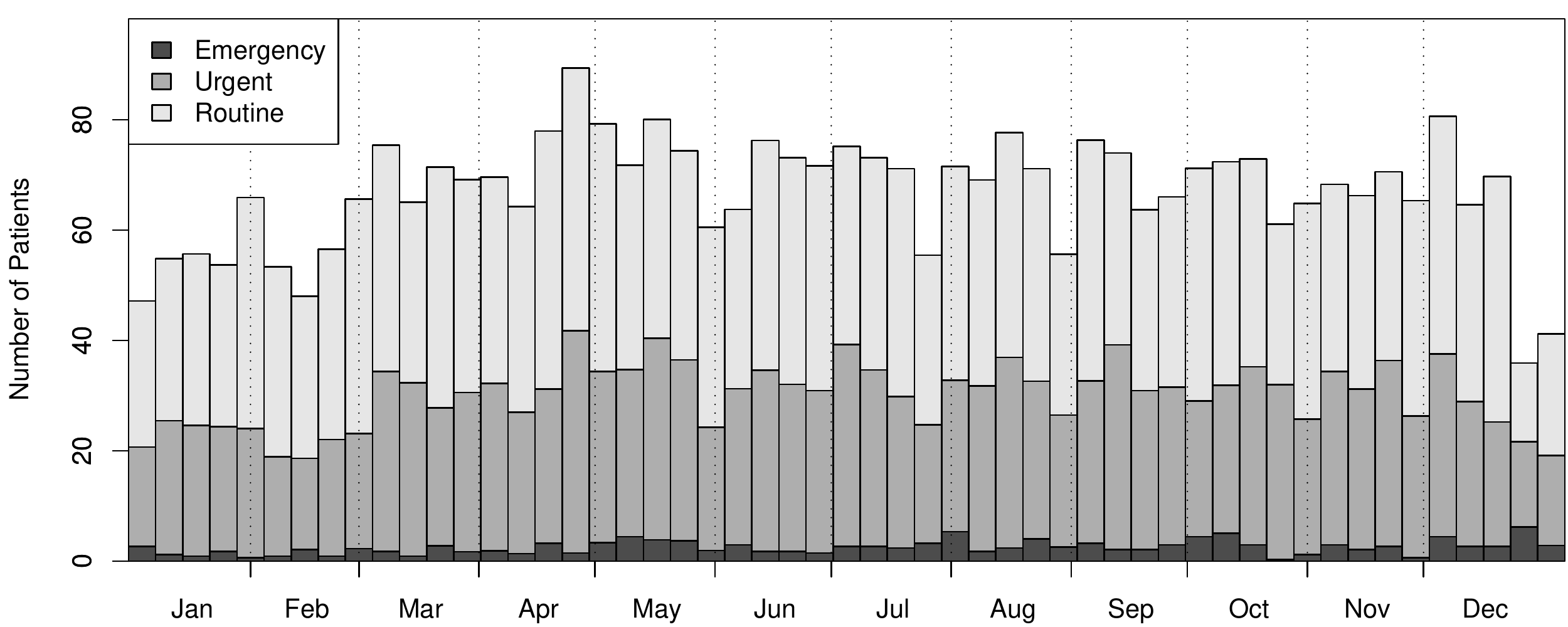}
  \caption{Average number of patients of each waiting list status per week during the year}
  \label{fig:patientsperweek}
\end{figure}

A seasonality is identified in the arrival of patients according to the week of the year, as can be seen in
\autoref{fig:patientsperweek}.
During the winter, the number of patients arriving each day is smaller than the year average. It slowly increases in
the next months, coming to a few peaks of patient arrivals in April and May. There is little variation in the next
months, ending with a steep drop in patient arrivals in the last two weeks of December. However, this variation is
slightly different for each waiting list status. The number of emergency patients has a smaller variation in the first
four months of the year, while the drop in the last two weeks of the year is not as steep for emergency and urgent as
it is for routine patients.



\begin{figure}[tb]
  \centering
  \includegraphics[width=.8\textwidth]{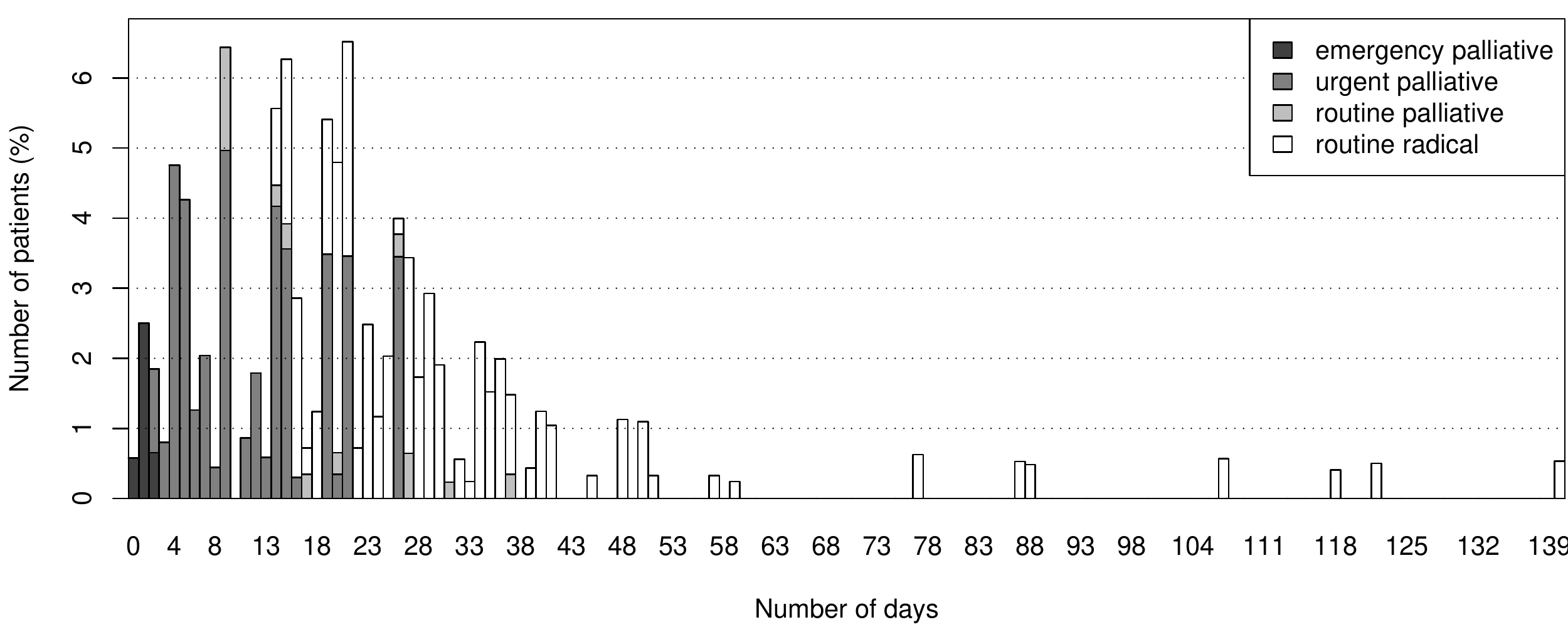}
  \caption{Histogram of the length of the time period between the decision to treat and release date given in days}
  \label{fig:pretreatment}
\end{figure}

\autoref{fig:pretreatment} shows a histogram of the length of the time period between the decision to treat and the
release date for each waiting list status/intent combination from the second data set. During this time, the patient
goes through the pre-treatment phase.

There seems to be a strong correlation between the length of this time period and the waiting list status/intent of
patients. Emergency patients have the shortest time period between the decision to treat and the release date of 1 day
in average. In contrast, urgent patients have a release date on average 11 days after the decision to treat has been
made, routine palliative patients an average of 18 days and routine radical patients an average of 33 days. The largest
values in \autoref{fig:pretreatment} can be explained by adjuvant patients, who first have a different treatment, such
as hormone therapy or chemotherapy, and then have adjuvant radiotherapy. For these cases, the breach date is calculated
from the date of the CT scan during pre-treatment instead of from the decision to treat.

It should also be noted that the differences presented in \autoref{fig:pretreatment} make it impossible for some
patients to meet all due dates. It is impossible for 17\% of emergency patients to meet their JCCO good practice of 1
day due to their release date being after this due date, but it is possible for all of them to meet the JCCO maximum
acceptable of 2 days. Around 94\% of non-emergency palliative patients cannot meet the JCCO good practice of 2 days,
and 23\% cannot meet the JCCO maximum acceptable of 14 days. For radical patients, the due dates are even harder to
meet, as 98\% of radical patients cannot meet the JCCO good practice of 14 days and 45\% cannot meet the maximum
acceptable of 28 days. In addition, 12\% of patients, all of which are routine and radical, cannot meet the breach date
of 31 days. This analysis can also give an approximation of the best possible values for the first three criteria.

\begin{figure}[tb]
  \centering
  \subfigure[Histogram of the number of sessions for each patient type]{
    \includegraphics[width=.47\textwidth]{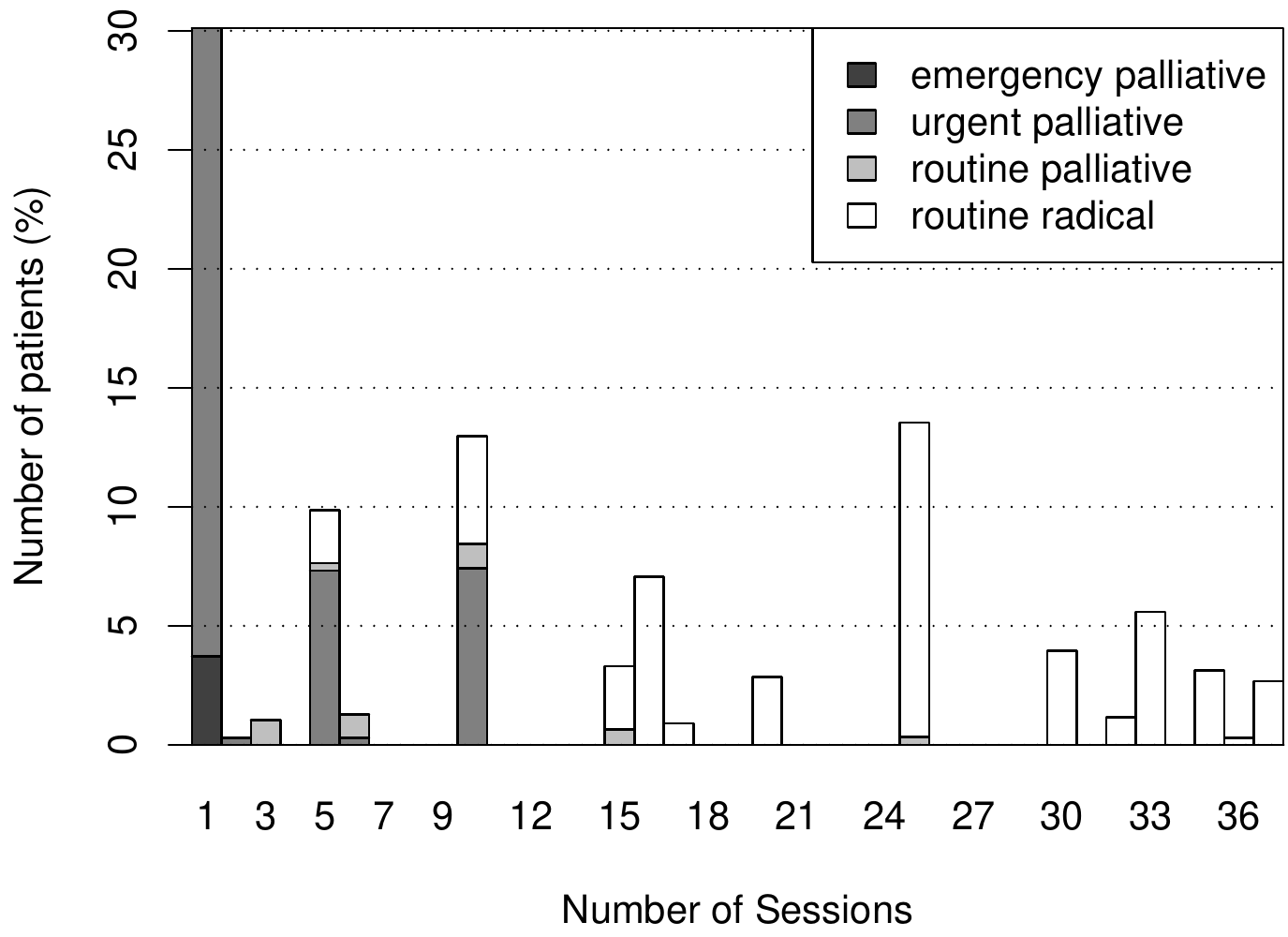}
    \label{fig:fractions}}
  \subfigure[Number of patients with each number of session days per week/sessions per day of each waiting list status]{
    \includegraphics[width=.47\textwidth]{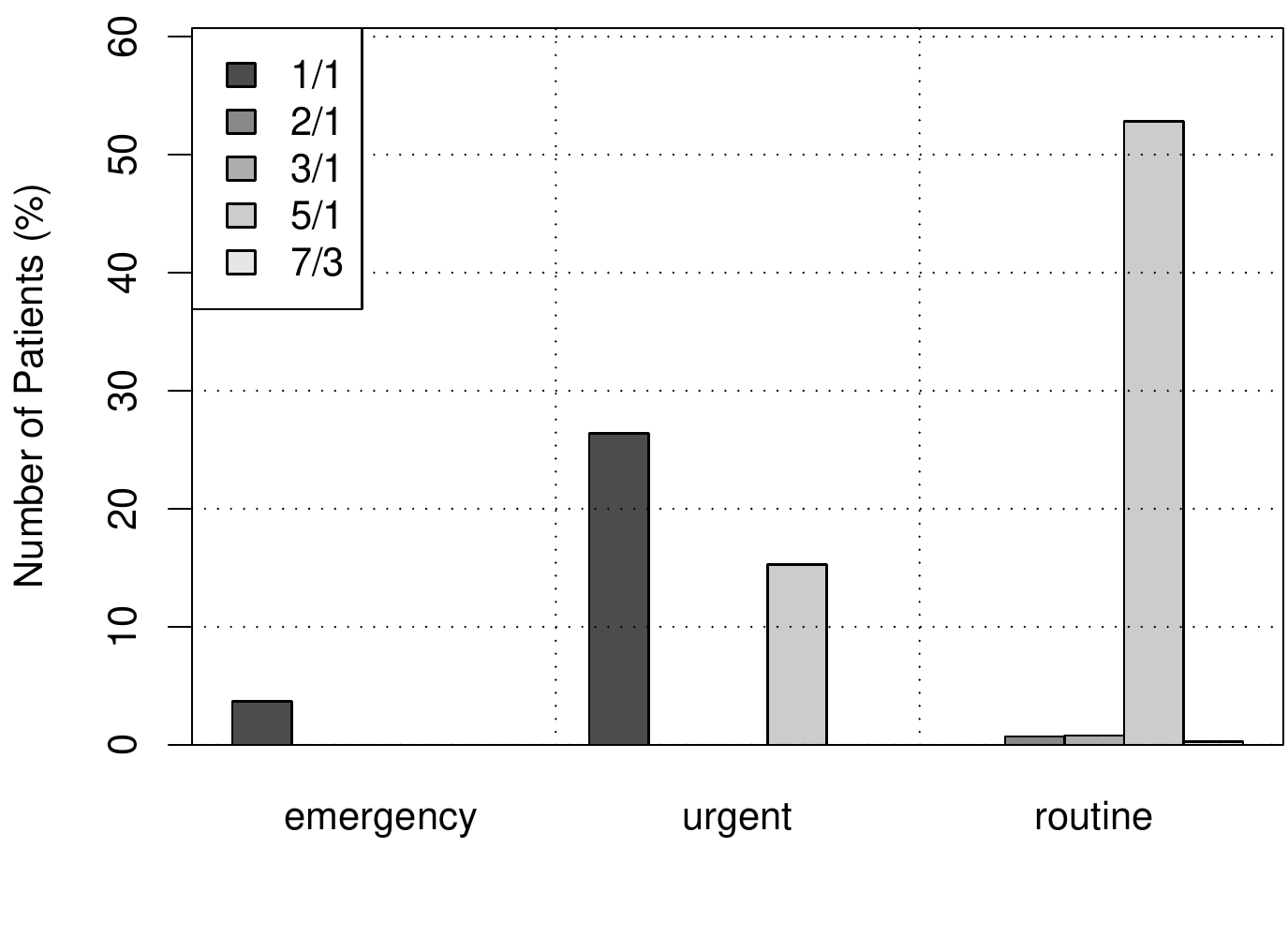}
    \label{fig:sessionsperday}}
  \caption{Frequency of patient types in each data set}
  \label{fig:sessiondays}
\end{figure}

Other important aspects of the data include the number of sessions of each patient and the number of session
days/sessions per day, shown in \autoref{fig:sessiondays}.

There seems to be a strong correlation between the number of sessions and the waiting list status/intent, as can be
seen in \autoref{fig:fractions}. All emergency, and around 63\% of urgent patients, have only one fraction. Routine
patients usually have a very high number of sessions, with an average of 21 sessions for each patient. Also, around
64\% of the patients who have more than 1 session have a number of fractions multiple of 5, showing a preference for
treatments which take a round number of weeks.

\autoref{fig:sessionsperday} shows how many patients of each waiting list status have each possible number of session
days per week/sessions per day. There also seems to be a correlation between these attributes and the waiting list
status of each patient. It is possible to see that the majority of patients (around 68\%) have 5 sessions per week and
1 session per day, where the exceptions are:
\begin{itemize}
  \item all emergency and most urgent patients who have 1 session day per week,
  \item a few routine patients who have 2 or 3 sessions per week,
  \item and a few routine patients classified as CHART, who have 3 sessions per day, 7 days per week for 12 consecutive
  days.
\end{itemize}


The oncology ward in the Nottingham University Hospitals NHS Trust, UK, currently has four linacs in total: 
\begin{itemize}
  \item 1 that emits low energy photon radiation (type $A$),
  \item 1 that emits low energy photon and electron radiation (type $B$),
  \item 2 that emit all three types of radiation (type $C$).
\end{itemize}
Linacs are available from 8:45 to 18:00 on Monday to Friday for weekday sessions and from 9:00 to 13:00 on Saturdays
and Sundays for weekend sessions.

In order to mimic the current scheduling policies in the hospital, a simplification is made. Patients who require only
low energy photon must be scheduled on linacs of type $A$, and patients who require electron and low energy photon or
only electron radiation must be scheduled on linacs of type $B$. Patients who require high energy photon can only be
scheduled on linacs of type $C$, regardless of requiring additional radiation types. This implies a simplification of
the problem, which is discussed later in this paper.

Two further simplifications are made in this work: random machine down times and patients not showing up for treatments
are not considered. However, the model can still be used as presented in both situations by adjusting each available
capacity $c_{ik}$ and by implementing a recovery protocol to reschedule sessions missed by patients. This protocol
could be based on integer linear programming or on other techniques.

In this work, all experiments are run using ILOG CPLEX 12.1, an optimisation software package, on a PC with an AMD
Opteron 2.2GHz CPU and 2GB of RAM under the Scientific Linux operating system.

\subsection{Experiments with Schedule Creation Day (SCD)}

Although using the model on its own on a daily basis can cause the earliest appointments to be always used first,
changes can be made to the scheduling policy to counteract this effect. Possible changes have been presented in
\cite{petrovic2008,petrovic2008a} involving the preferred date for patients to start, days when schedules are created
and machine reservation. Two of those approaches are also analysed in this paper by using the proposed model with a
standard ILP solver in place of the constructive algorithm previously presented. The first approach introduces a
parameter called \emph{schedule creation day} (SCD) to specify days of the week for each patient when a schedule can be
created for patients of a given waiting list status. If a patient has a decision to treat made on a date when the
creation of a schedule is not allowed, the schedule is created on the first following allowed day.

The intention is to investigate whether the accumulation of patients to be scheduled will lead to better schedules.
Obviously, the search space becomes larger and it may lead to solutions of higher quality. The SCD values considered
are 5 (every weekday), 3 (on Mondays, Wednesdays and Fridays), 2 (on Tuesdays and Fridays) and 1 (only on Fridays) for
urgent and routine patients and fixed as 7 (every day) for emergency patients.

The model presented is dependent on the value of the scheduling horizon $T$, which must be supplied as input and must
be large enough to accommodate all patients. To calculate this value, the constructive algorithm presented in
\cite{petrovic2008a} is used. The proposed algorithm sorts the patients to be scheduled according to their waiting list
status, breach date, JCCO maximum acceptable target and number of sessions. Following that order, each patient is
scheduled on the earliest day possible. The value of $T$ is taken from the date of the last session in this solution
and increased by 14 days in order to augment the search space for CPLEX. Since $T$ is taken from a feasible solution,
it is impossible for the instance to have no feasible solutions. The solution found by the constructive approach is
also used as starting point by CPLEX in order to increase its execution speed.

%

The schedule of patients who must be scheduled on linacs of one type has no influence on the schedule of patients who
must be scheduled on linacs of other types. Therefore, it is safe to split the problem into three sub-problems in order
to speed up the process of finding a schedule each day. Each sub-problem is composed of linacs of one type and all
patients who must be scheduled on linacs of that type. Each sub-problem is solved individually and the schedules of
patients are combined to form a complete schedule. With this simplification, linac eligibility constraints are no
longer necessary and can be removed. However, they are kept in order to avoid loss of generality in the model. The time
limit of 10 minutes is equally divided amongst the three sub-problems.

Each configuration is run on the 33 instances described in \autoref{sec:experiments}. When running a number of
experiments with randomly generated data, it is possible that one configuration achieves a better average result than
the other configurations simply by chance. In order to determine whether or not the means of the criteria values of two
sets of experiments are significantly different, the Mann-Whitney U (MWW) test is used on experiment values re-sampled
from a bootstrap approximation \citep{leger1992}. The MWW test \citep{mann1947} is able to determine if there is
significant statistical evidence that one configuration achieves better results than another configuration.
Bootstrapping is a computationally intensive technique based on data re-sampling. By using bootstrapping, it is
possible to perform valid statistical tests without making unrealistic or unverifiable assumptions about the criteria
values, like their distribution or variance \citep{efron1979,yuan2007}.

The MWW tests are run for each pair of configurations with an overall confidence of 90\% and the bootstrap is set to
1000 replications. Results for each criterion are shown in \autoref{tab:scd}, where ``Breach'' is the percentage of
patients who miss their breach date, ``JMax'' and ``JGood'' are the weighted percentage of patients who miss their JCCO
maximum acceptable and good practice targets, respectively, and ``Waiting'' is the average weighted squared waiting
time per patient.
The values in bold are the ones where there was no significant statistical evidence of any of the values found by the
other configurations being better for that criterion. This form of evaluation is also used in the subsequent sections.

\begin{table*}[tb]
  \centering
  \begin{tabular}{cc|cccc}
    \hline
    \multicolumn{2}{c|}{SCD} & Breach (\%) & JMax (\%) & JGood (\%) & Waiting \\
    \cline{1-2}
    Urgent & Routine         &             &           &            &         \\
    \hline \hline
5 & 5 & 34.98          & 47.67          & 86.59          & 1,670          \\
5 & 3 & 34.79          & 47.49          & 86.57          & 1,649          \\
5 & 2 & 34.89          & 47.56          & \textbf{86.51} & 1,670          \\
5 & 1 & 34.13          & 47.22          & \textbf{86.47} & \textbf{1,623} \\
3 & 5 & 35.08          & 47.58          & 86.59          & 1,682          \\
3 & 3 & 34.74          & 47.39          & \textbf{86.57} & 1,650          \\
3 & 2 & 35.01          & 47.48          & \textbf{86.53} & 1,678          \\
3 & 1 & 34.04          & \textbf{47.13} & \textbf{86.47} & \textbf{1,611} \\
2 & 5 & 34.92          & 47.53          & \textbf{86.54} & 1,665          \\
2 & 3 & 34.85          & 47.42          & \textbf{86.52} & 1,663          \\
2 & 2 & 34.68          & 47.40          & \textbf{86.49} & 1,657          \\
2 & 1 & \textbf{33.95} & \textbf{47.06} & \textbf{86.41} & \textbf{1,624} \\
1 & 5 & 35.37          & 47.34          & 89.05          & 1,680          \\
1 & 3 & 35.22          & 47.22          & 89.03          & 1,678          \\
1 & 2 & 34.92          & \textbf{47.24} & 89.03          & 1,668          \\
1 & 1 & \textbf{33.77} & \textbf{46.89} & 88.99          & \textbf{1,624} \\
    \hline
  \end{tabular}
  \caption{Results obtained varying the frequency of creating schedules for patients of each waiting list status.}
  \label{tab:scd}
\end{table*}

In general, good results for the breach date are found when routine patients are scheduled once a week, with the best
of these being when urgent patients are scheduled either once or twice per week. As the breach date is the least
restrictive target (it is the largest target date), it is possible to achieve better results by slightly delaying the
creation of schedules for patients, so that schedules are created only once a week to increase the search space.

For the JCCO maximum acceptable target, the best results are found when creating a schedule for urgent patients between
one and three times per week and for routine patients once per week, or for urgent patients once per week and for
routine patients twice a week. For a large portion of urgent patients, it is impossible to meet this target due to a
large time interval between the decision to treat and the release date. This is likely why the frequency of creating
schedules for urgent patients does not have such a large influence on the quality of this criterion.

When considering the JCCO good practice target, the best results were obtained either creating schedules for urgent
patients five times per week and for routine patients twice or less, or for urgent patients three times per week and
for routine three times or less, or for urgent patients twice a week. It is possible to see a pattern with the value of
the SCD parameter and the restrictiveness of each target date. The more restrictive the target date is, the more
frequently the creation of schedules achieve better results. This makes sense since the more restrictive the target
date is, the more likely it is that the patient's release date will have already passed when the schedule is created
for that patient if the schedule for them is created with a low frequency.

For the squared waiting time, the best results are found when routine patients are scheduled once a week, similarly to
the breach date. The authors believe that the configurations that achieve the best results for the breach date also
achieve the best results for the squared waiting time because both criteria give a much greater penalty to patients who
have large waiting times than patients with shorter waiting times.

When using a lexicographical approach, it is possible that the first criteria constrain the search space in a way that
there is no room for improvement for the other criterion. However, this is rarely the case in the problem investigated
in this work, as the criteria presented are increasingly more restrictive. Apart from situations where the criteria
value is the same for all feasible solutions (e.g. a day when all patients have their release date after their JCCO
good practice target), objectives are improved in around 56\% of the times.


The values for the SCD parameter of 2/1 for urgent/routine patients achieve the best results for all criteria and are,
therefore, used in the remaining experiments.

\subsection{Experiments with Maximum Number of Days in Advance (MNDA)}

Two important dates in radiotherapy treatment scheduling are the decision to treat date and the release date. Since
these two dates can be very far apart, one possibility of achieving a better schedule is to not create a schedule for
the patients immediately when they arrive in the radiotherapy centre, but to wait for the release date to become
closer, i.e. towards the end of their pre-treatment phase. This might give better chances of good quality schedules for
patients that will arrive in the near future, while still obtaining good performance for current patients.

The maximum number of days in advance (\emph{MNDA}) parameter is introduced to limit the creation of schedules based on
the patient's release date. The values used in the experiments for the MNDA parameter of urgent and routine are:
\begin{itemize}
  \item $\infty$ (infinity) - the schedule is created as soon as the patient arrives,
  \item 21, 14, 7 - the schedule is created when the release date is within 21, 14 or 7 days, respectively,
  \item and 0 - the schedule is created on the release date or afterwards.
\end{itemize}
For emergency patients, the MNDA value is fixed at $\infty$. Results are presented in \autoref{tab:mnda}.

\begin{table*}[tb]
  \centering
  \begin{tabular}{cc|cccc}
    \hline
    \multicolumn{2}{c|}{MNDA} & Breach (\%) & JMax (\%) & JGood (\%) & Waiting \\
    \cline{1-2}
    Urgent & Routine          &             &           &            &         \\
    \hline \hline
$\infty$ & $\infty$ & 33.95          & 47.06          & 86.41          & 1,624          \\
$\infty$ & 21       & 29.84          & 45.55          & 86.11          & 1,416          \\
$\infty$ & 14       & 27.72          & 44.04          & 85.95          & 1,293          \\
$\infty$ &  7       & \textbf{26.81} & 40.92          & 85.75          & 1,155          \\
$\infty$ &  0       & 31.46          & 40.25          & \textbf{85.60} & \textbf{1,122} \\
21       & $\infty$ & 33.96          & 47.04          & 86.41          & 1,624          \\
21       & 21       & 29.85          & 45.53          & 86.11          & 1,416          \\
21       & 14       & 27.73          & 44.02          & 85.94          & 1,293          \\
21       &  7       & \textbf{26.82} & 40.90          & 85.75          & 1,155          \\
21       &  0       & 31.47          & 40.23          & \textbf{85.60} & \textbf{1,122} \\
14       & $\infty$ & 34.00          & 46.97          & 86.40          & 1,628          \\
14       & 21       & 30.00          & 45.42          & 86.10          & 1,419          \\
14       & 14       & 27.90          & 43.87          & 85.92          & 1,307          \\
14       &  7       & \textbf{26.84} & 40.76          & 85.73          & 1,155          \\
14       &  0       & 31.51          & \textbf{40.01} & \textbf{85.60} & \textbf{1,136} \\
 7       & $\infty$ & 34.38          & 46.85          & 86.38          & 1,684          \\
 7       & 21       & 30.82          & 45.38          & 86.08          & 1,500          \\
 7       & 14       & 28.53          & 44.14          & 85.90          & 1,346          \\
 7       &  7       & 27.30          & 41.16          & \textbf{85.65} & 1,191          \\
 7       &  0       & 31.64          & 40.48          & \textbf{85.57} & 1,169          \\
 0       & $\infty$ & 35.44          & 47.66          & 88.73          & 1,772          \\
 0       & 21       & 32.18          & 46.16          & 88.45          & 1,587          \\
 0       & 14       & 29.85          & 45.49          & 88.36          & 1,421          \\
 0       &  7       & 28.05          & 43.69          & 88.21          & 1,267          \\
 0       &  0       & 32.30          & 43.61          & 88.31          & 1,225          \\
    \hline
  \end{tabular}
  \caption{Results obtained varying the antecedence with which schedules are created.}
  \label{tab:mnda}
\end{table*}

For the breach date, the best results are obtained when creating schedules for urgent patients when their release date
is within 14 or more days and for routine patients when their release data is within 7 days. It is possible to see a
large difference in the number of late patients between creating schedules when the release date is within 7 days and
on the release date or afterwards (MNDA values 7 and 0, respectively). This can be explained by the fact that when
creating schedules once per week on the release date or after it (MNDA value 0), there might be a patient whose
schedule is created only after the breach date, while if the schedule had been created when the release date was within
7 or more days, it would have been created before the breach date, thus not violating it.

For the JCCO maximum acceptable target date, which is slightly more restrictive than the breach date, the best results
are obtained by scheduling the urgent patients when their release date is within two weeks and routine patients either
on their release date or after. This way, it is possible to give a higher priority for creating schedules for the
urgent patients, without compromising the schedules of emergency patients.

The JCCO good practice is the most restrictive target date. The best results for this criterion are found when creating
schedules for urgent patients when their release date is within 7 or more days and routine patients on their release
date or afterwards, or when creating schedules for both types of patients when their release date is within 7 days.

The values of the squared waiting time criterion vary similarly to the values of the JCCO good practice objective
function. Good results are achieved when creating schedules for urgent patients within 14 or more days of their release
date and for routine patients on their release date or after. These values lead to schedules where only few patients
have very large waiting times.

Using the parameters values of 14/0 for urgent/routine patients achieves the best results for 3 of the 4 examined
criteria. However, the authors recommend using the values $\infty$/7 instead, as they achieve the best value for the
breach criterion, and values very close to the best values obtained for the remaining criteria. In comparison, using
the values 14/0 achieves the best values for the JCCO maximum acceptable, good practice and squared waiting time, but
the values achieved for the breach criteria are considerably worse than the best found.

\section{Conclusions and Future Work}
\label{sec:conclusion}

This research investigates the problem of scheduling the treatments of radiotherapy patients in the Nottingham
University Hospitals NHS Trust, UK. A new integer linear programming model with four optimisation criteria is
formulated and data for the problem is generated based on real-world data from the hospital.

Throughout this paper, different policies for scheduling treatments for radiotherapy patients were experimented with.
It was demonstrated to be possible using an efficient optimisation tool, such as CPLEX, for finding a schedule of good
quality. It might be worth waiting for an accumulation of patients in order to increase the search space and find a
schedule of good quality for a larger number of patients. This can be done, for instance, by using this tool once per
week for creating schedules for patients.

The experiments also suggest that it is better to not create schedules for patients immediately when they arrive, but
to wait for their release date to become closer. It is understood that patients prefer to know their treatment schedule
as soon as possible, but in view of the possible improvements in the quality of their schedule, the authors recommend
waiting for their release date to become closer before creating their schedule. The best results in the experiments
were achieved by creating schedules for routine patients only when their release date was within 7 days.

Future work includes the investigation of look-ahead techniques to try to anticipate how many patients of each category
might arrive in the succeeding days. Criteria to evaluate the robustness of a solution with regard to future patients
can also be investigated as well as the implementation of rescheduling policies.

So far, we have considered allocation of treatment days to patients. More constraints will be included to realistically
capture the real-world radiotherapy scheduling problem, such as considering the random machine down times, patient
preference for being treated in morning or afternoon sessions, requirements for transportation to and from the
hospital, etc.

The implementation of a graphical user interface for every day use in the hospital is currently in development, and it
will be placed for use in the hospital at the end of the project.

\section*{Acknowledgements}

The authors would like to thank the Nottingham University Hospitals NHS Trust, UK, and the Engineering and Physics
Sciences Research Council (EPSRC), UK for supporting this research (Ref No. EP/C549511/1).

\bibliography{bibliography}

\end{document}